\documentstyle[12pt,a4,psfig,epsfig]{article}
\begin{document}
%\begin{titlepage}
\begin{flushright}
TAUP 2750-03
\end{flushright}
% epsfig package included for placing EPS figures in the text
%------------------------------------------------------

%%%%%%%%%%%%%%%%%%%%%%%%%%%%%%%%%%%%%%%%%%%%%%%%%%
%                                                %
%    BEGINNING OF TEXT WEB VERSION                          %
%                                                %
%%%%%%%%%%%%%%%%%%%%%%%%%%%%%%%%%%%%%%%%%%%%%%%%%%

%\listoffigures
\begin{center}
{\bf OPEN QUESTIONS RELATED TO BOSE-EINSTEIN\\ CORRELATIONS IN
%questions related to Bose-Einstein correlations in
$e^+e^- \to$ HADRONS
\footnote{Presented at the ISMD2003, 5$-$11 September 2003, Krakow, Poland}\\
%you can use '\\' to break lines
}
\vspace{1cm}

{\large\bf Gideon Alexander}\\
\vspace{2mm}

{\large School of Physics and Astronomy\\ Tel-Aviv University,
Tel-Aviv, Israel}
\end{center}
%\maketitle

\begin{abstract}
Questions concerning the Bose-Einstein (BEC) and Fermi-Dirac (FDC)  
correlations of hadrons produced in $e^+e^-$ collisions are discussed.
Among them the emitter dimension $r$ as a function of $\sqrt{s_{ee}}$ 
and the hadron
mass, the extension of the BEC by including isospin invariance and
the proposed relation between $r$ and the
inter-atomic separation in Bose condensates.
\end{abstract}
%\PACS{13.85.Hd, 03.75.Fi, 05.30 Jp, 13.65 +i}
%\end{titlepage}
\section{Introduction}
The present report covers a
few of the open questions which are related to
the Bose-Einstein and Fermi-Dirac Correlations (BEC and FDC)
of hadrons in $e^+e^-$ annihilations.
The topic which are dealt here include the dependence of the 
hadron emitter dimension 
on the $\sqrt{s_{ee}}$ and on the mass of the hadron; the 
interpretation  
of the 2-dimensional BEC analysis results obtained  
in annihilations as compared to those found in heavy ion collisions and
finally the status of the so called Generalised BEC. The 
issue whether one may relate some features of the BEC at high
energies with those present in Bose condensates 
is also briefly discussed.

\section{Dependence of the dimension $r$ on the $\sqrt{s_{ee}}$}
\begin{figure}[h]
\centering{\epsfig{file=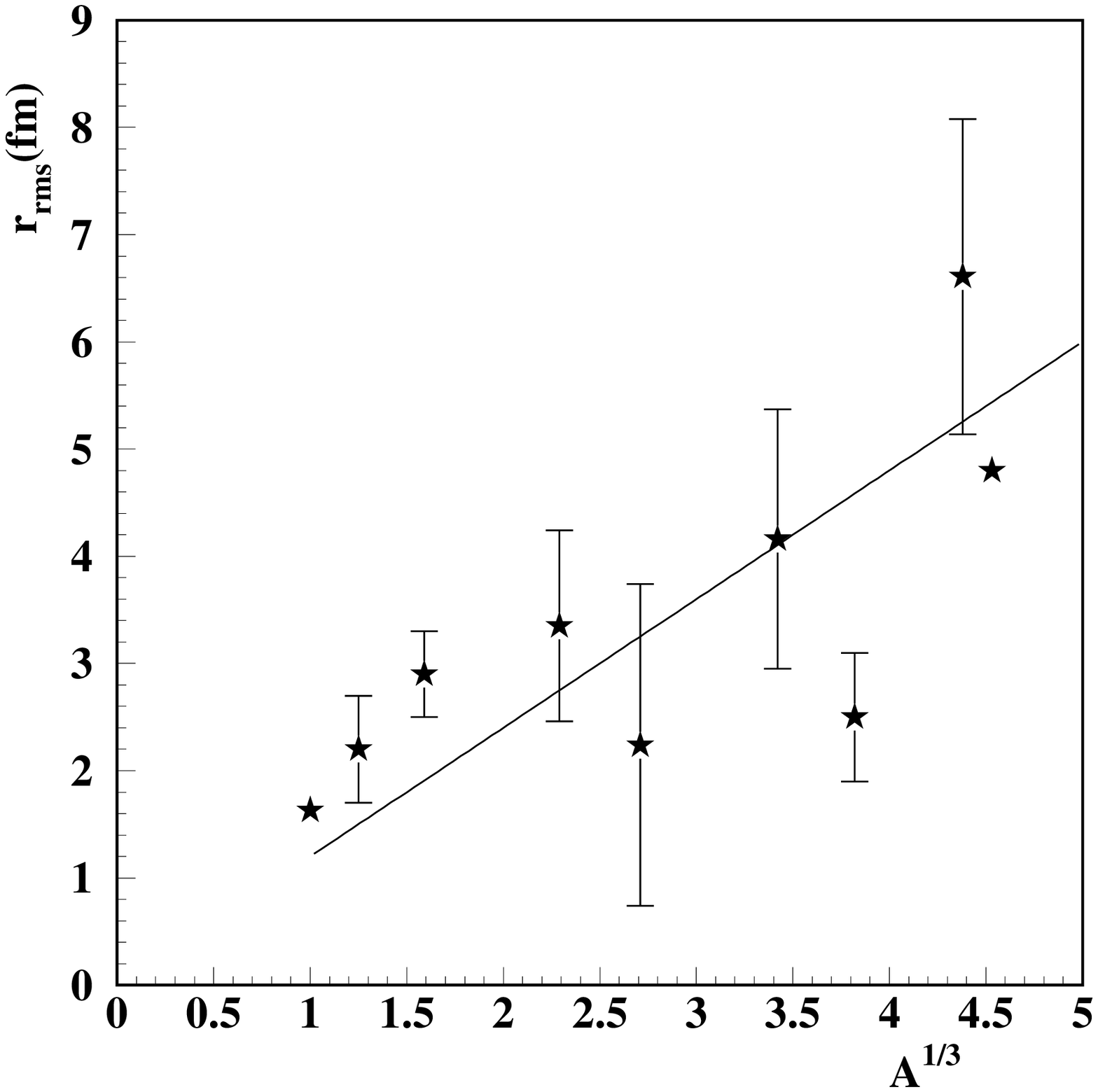,height=4.5cm,
bbllx=32pt,bblly=156pt,bburx=532pt,bbury=664pt,clip=
}\ \ \ \
\ \epsfig{file=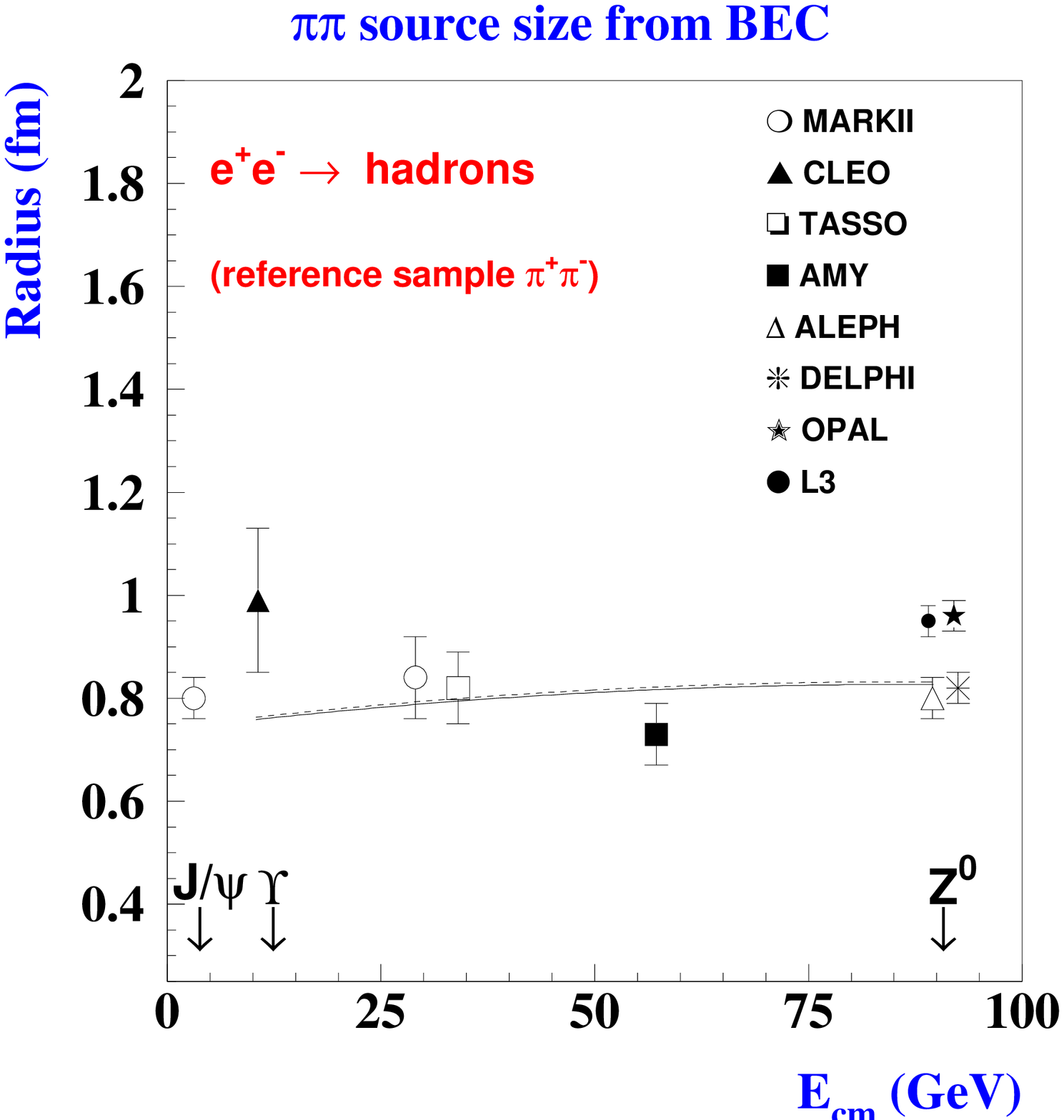,height=4.8cm}}
%bbllx=32pt,bblly=156pt,bburx=532pt,bbury=664pt,clip=}}
\caption{\small
Left: 
$r_{\pi\pi}$ versus A$^{1/3}$ found 
in heavy ion
collisions. The line represents $r=1.2A^{1/3}$ fm.
Right: $r_{\pi\pi}$ versus $\sqrt{s_{ee}}$ found in $e^+e^-$ annihilations.
The line is the expectation of a hadron jet source model 
(see text).}
\label{fig_compr}
\end{figure}

As is well known the Bose-Einstein
interferometry method rest on the fact that
the density of identical boson pairs
is enhanced when they are emitted near in momentum and phase space.
Assuming the emitter to be a sphere of a Gaussian
distribution, the correlation of pairs of identical bosons 
can be parametrised  
by $C_2(Q)=1+\lambda_2e^{-r^2Q^2}$ where $Q^2=-(p_1-p_2)^2$ and 
$p_1$ and $p_2$ are the four momenta of the two bosons.
An early
compilation of the $r_{\pi\pi}$ values measured in heavy nuclei
(AA) reactions is seen in Fig. \ref{fig_compr} 
to be rather well described by $r=1.2A^{1/3}$ fm,
the known nucleus radius dependence on the
atomic number A.  
This behaviour is taken to be the result of the increase 
of pion sources as A gets larger. 
As for the 
$r_{\pi\pi}$ values obtained from the $e^+e^-$ annihilation analyses, they 
are seen in Fig. \ref{fig_compr} to be essentially independent
of $\sqrt{s_{ee}}$ as it is generally accepted that 
the small deviations 
from a constant value are due to
the non-uniformity of the chosen
experimental procedures. Since e.g. the  
average hadron multiplicity increases with energy
this flat $r$ distribution poses a theoretical challenge.\\

\noindent
In an effort to understand the origin of this flat $r$ distribution
we adopt here, for the $e^+e^-$ annihilations,
the hadron jets as the pion sources 
and further assume that only pion pairs emerging from the same jet
may be correlated. Now the effect of multi-hadron sources 
on particles' correlation strength 
has been studied by several authors \cite{review} 
in the frame work of the factorial cumulants moments $K_q$ where 
$q=2$ in the case of pion-pair correlations. 
Specifically, it
has been shown that the cumulant is diluted by a factor $D_q$, 
so that $K^D_q=D_q\cdot K^{D=1}_q$ where $D_q$ is defined as
$D_q=N^{same}_q/(N^{same}_q+N^{different}_q)$. Here $N^{same}_q$
and $N^{different}_q$
denote respectively the number of $q$ pions combinations 
coming from the same and
from different hadron jets. Now for $q=2$ there exists the relation 
$C_2(Q)=1+\lambda_2 e^{-Q^2r^2}=1+K_2$ which
can further be extended to
include the dilution factor $D(=D_2)$ so that
$$C_2^D(Q)\ =\ 1 + K^D_2\ =\ 1 + DK_2^{D=1}\ =\ 1 + \lambda_D D
e^{-Q^2r^2_{D=1}}\ =\ 1 + \lambda_D e^{-Q^2r^2_D}\ .$$
The dependence of $Q$ in this expression is then removed by
integrating the quantity $C_2(Q)-1$ over $Q$ from zero to infinity to 
finally obtain
$$r_D\ =\ \frac{1}{D}\frac{\lambda_D}{\lambda_{D=1}}r_{D=1}\ .$$
To apply this formula to the $e^+e^- \to hadrons$ reaction there is the
need to evaluate the dilution factor $D$
as a function of $\sqrt{s_{ee}}$. As is well known the determination of
the number of hadron jets in a particle reaction does depend on the
chosen jet
identification
method \cite{bot}. Here for simplicity we will 
assume that the $e^+e^-$ annihilation proceeds via three hadron jets,
two coming from the quark and anti-quark pair and the third from a
radiated gluon\footnote{Even at the Z$^0$ the fraction of events with more
than three hadron jets is very small if one uses e.g. $y_{cut}>0.01$ in the 
Durham jet identification scheme \cite{jet}.}. 
The total $\sqrt{s_{ee}}$ is then divided in the
following way. The first quark jet gets on the average the energy 
$\sqrt{s_{ee}}/2$ and thus the second quark jet has the
energy of $\sqrt{s_{ee}}/2-E_{gluon}$. The final number of outgoing 
charged
pions assigned to each of the three jets is determined from the 
known total average charged hadron multiplicity 
dependence on $\sqrt{s_{ee}}$ and the known average charged multiplicity of
quarks and gluon jets parametrised as a function of their jet energy
\cite{bot}. 
From this pion division the dilution factor $D_2$ is estimated and
the resulting $r$ dependence on $\sqrt{s_{ee}}$ is shown 
in Fig. \ref{fig_compr}
by the continuous line which is normalised at 40 GeV
while setting $\lambda_D =\lambda_{D=1}$.
As seen, the $r$ behaviour extracted from this simplified
model, which rises by only $\sim$3.5$\%$ from 10 to 90
GeV CM energy, already describes nicely 
experimental situation.  
\section{ The $r$ dependence on the hadron mass}
\begin{figure}[h]
\centering{\epsfig{file=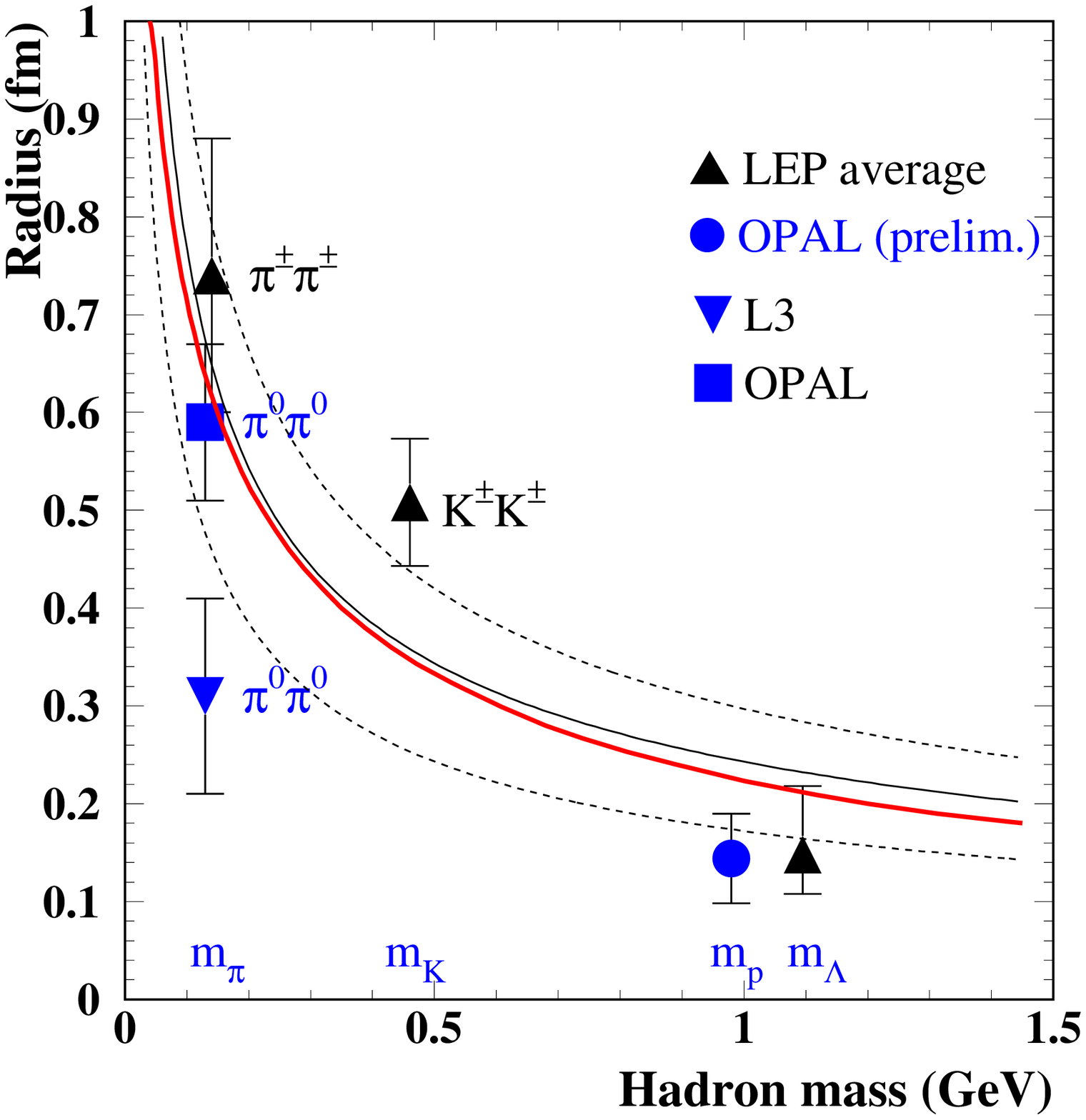,height=5.5cm}
%bbllx=32pt,bblly=156pt,bburx=532pt,bbury=664pt,clip=}
\ \ \ \
\ \epsfig{file=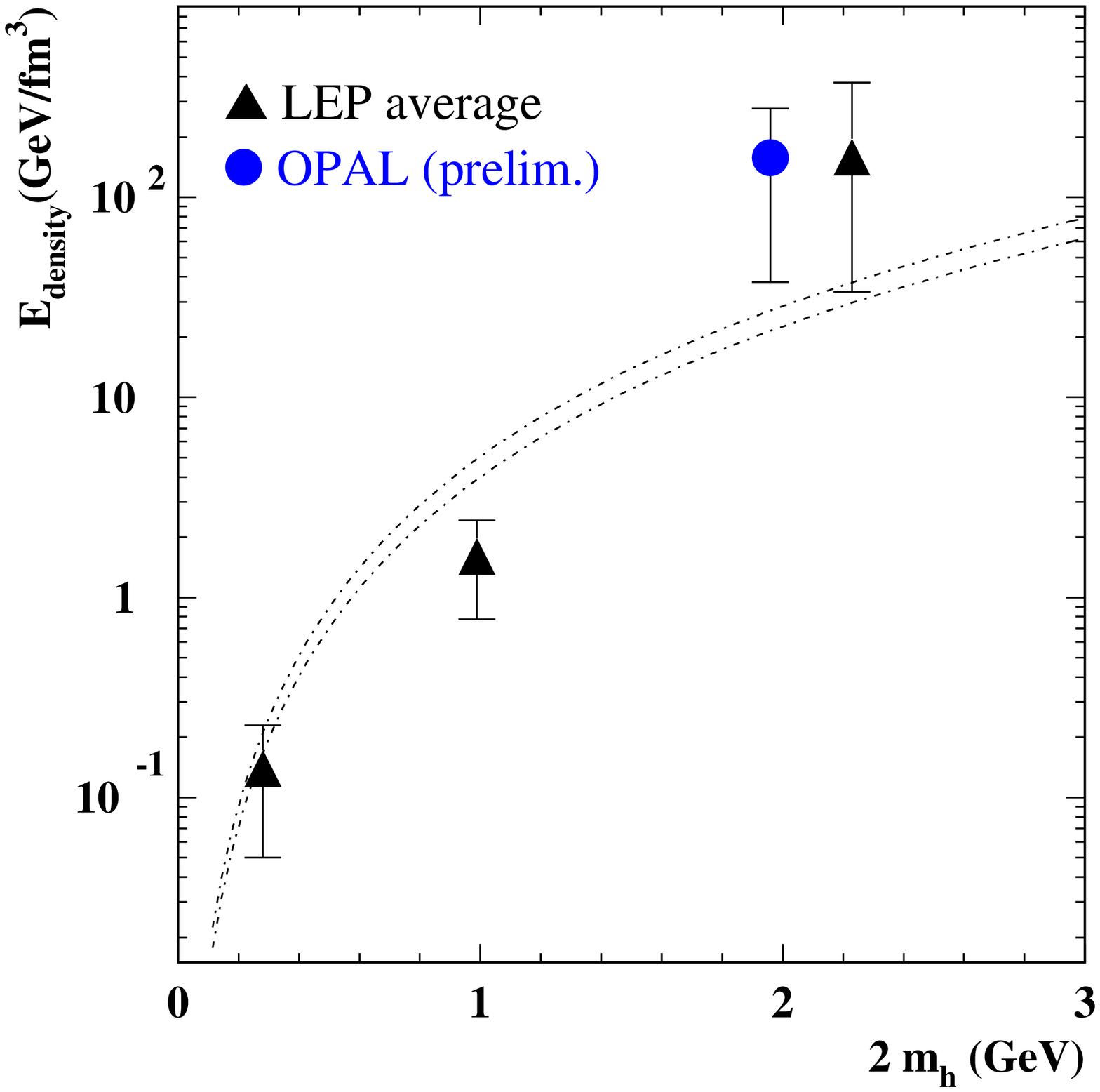,height=5.5cm}}
%bbllx=32pt,bblly=156pt,bburx=532pt,bbury=664pt,clip=}}
\caption{\small Left: The $r$ dependence on the hadron mass
in hadronic Z$^0$ decays. Right: The minimum energy density of the
hadron emitter \cite{review} versus the hadron mass.
}
\label{fig_mass}
\end{figure}

The very high statistics of the Z$^0$ hadronic decay events accumulated by
the LEP experiments and the extension
of the BEC analysis to baryon pairs via the 
FDC method opened the way to study the emitter 
dimension $r_{\pi\pi}$ as a function
of the hadron mass, from pions to $\Lambda$ baryons.
The results of these analyses are shown in Fig. \ref{fig_mass} 
together with two
recent, L3 and OPAL, measured $r_{\pi^0\pi^0}$
values which are seen to be inconsistent and therefore are
presently disregarded.
The emitter dimension $r(m)$ 
is seen 
to decrease with the hadron mass, from 
$\sim$0.75 fm at $m_{\pi}$ to $\sim$0.15 fm at $m_{\Lambda}$.
It has further been shown \cite{acl} that this 
behaviour can be derived from the Heisenberg uncertainty
relations which yield 
\begin{equation} 
r(m)=c\sqrt{\Delta t ~\hbar}/\sqrt{m}\ . 
\label{eq_acl}
\end{equation}
Choosing $\Delta  t = 10^{-24}$ second, as a    
representative time scale of the the strong interaction sector, $r(m)$
is shown in Fig. \ref{fig_mass} by the continuous line. 
An almost identical expectation for $r(m)$ is also 
forthcoming from the Local
Parton Hadron Duality and a general QCD potential \cite{acl}. 
On the other hand, the decrease of $r$ with the hadron mass 
poses a challenge to a large variety of hadronisation
models, including the Lund one, which expect $r$ to increase with
the hadron mass. In particular at present,
there is no satisfactory explanation for the very small  
$r_{\Lambda}$ value.\\ 
In heavy ion collisions  
part of the emitter survives after the emission of
the identical bosons used in the BEC analysis. 
This is not the case in $e^+e^-$ annihilation where he emitter 
collapses with the emission of its hadrons used for the BEC analysis. 
Consequently one may try and estimate an approximate
minimum energy density of the emitter of the two hadrons by the relation 
$E_{density}=2m_h/V=6m_h/(4\pi r^3)$ which is shown in Fig. \ref{fig_mass}.
Whereas the energy density for the pion and kaon emitters still
lies within a reasonable range, the energy densities of the 
protons and $\Lambda$'s emitters are lying in the vicinity of 100 GeV/fm$^3$, 
way above the values anticipated from the current models for hadron
production.
\section{The 2-dimensional BEC analyses}
\begin{figure}[h]
\centering{\epsfig{file=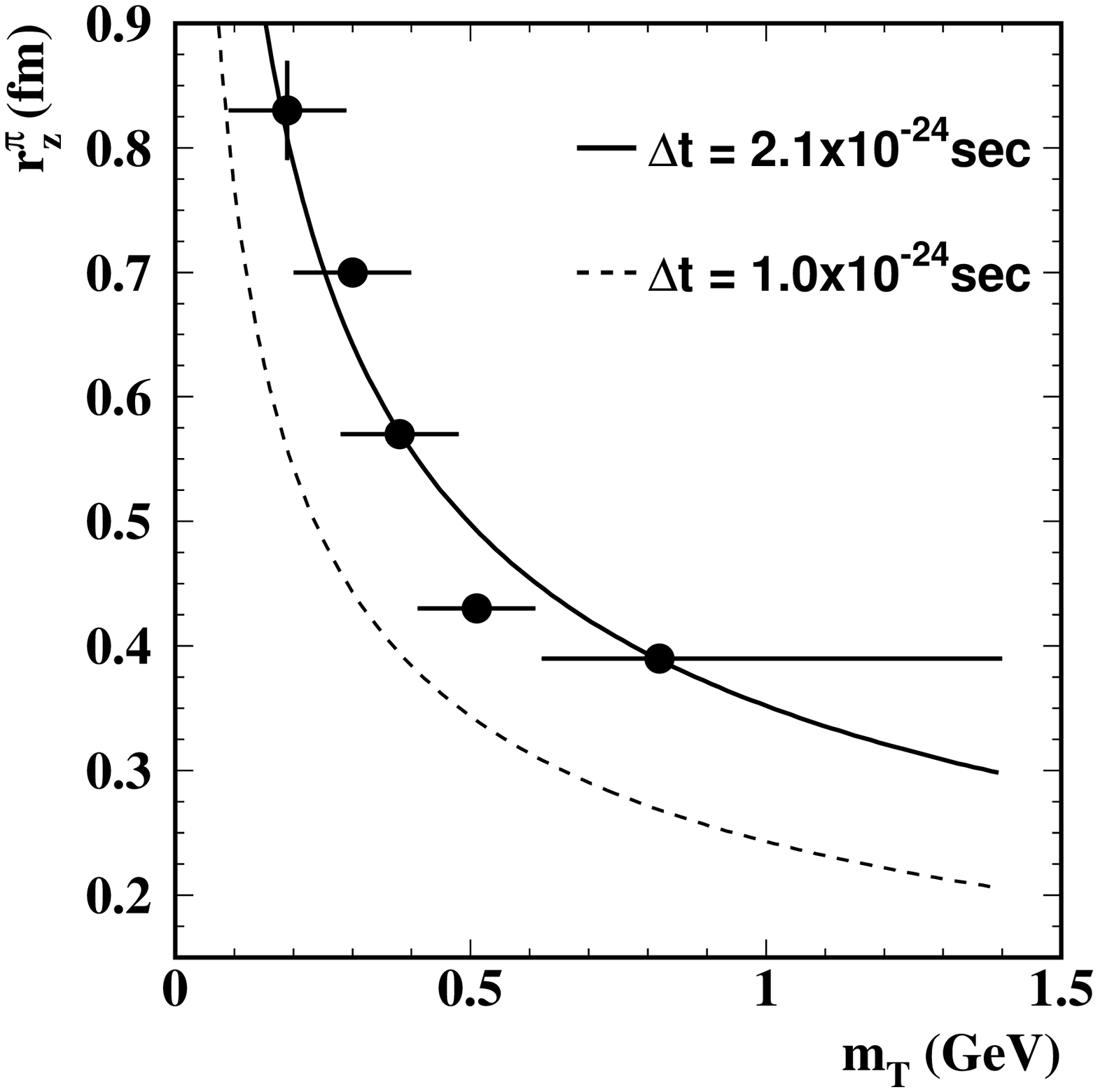,height=5.cm}
\ \ \ 
\epsfig{file=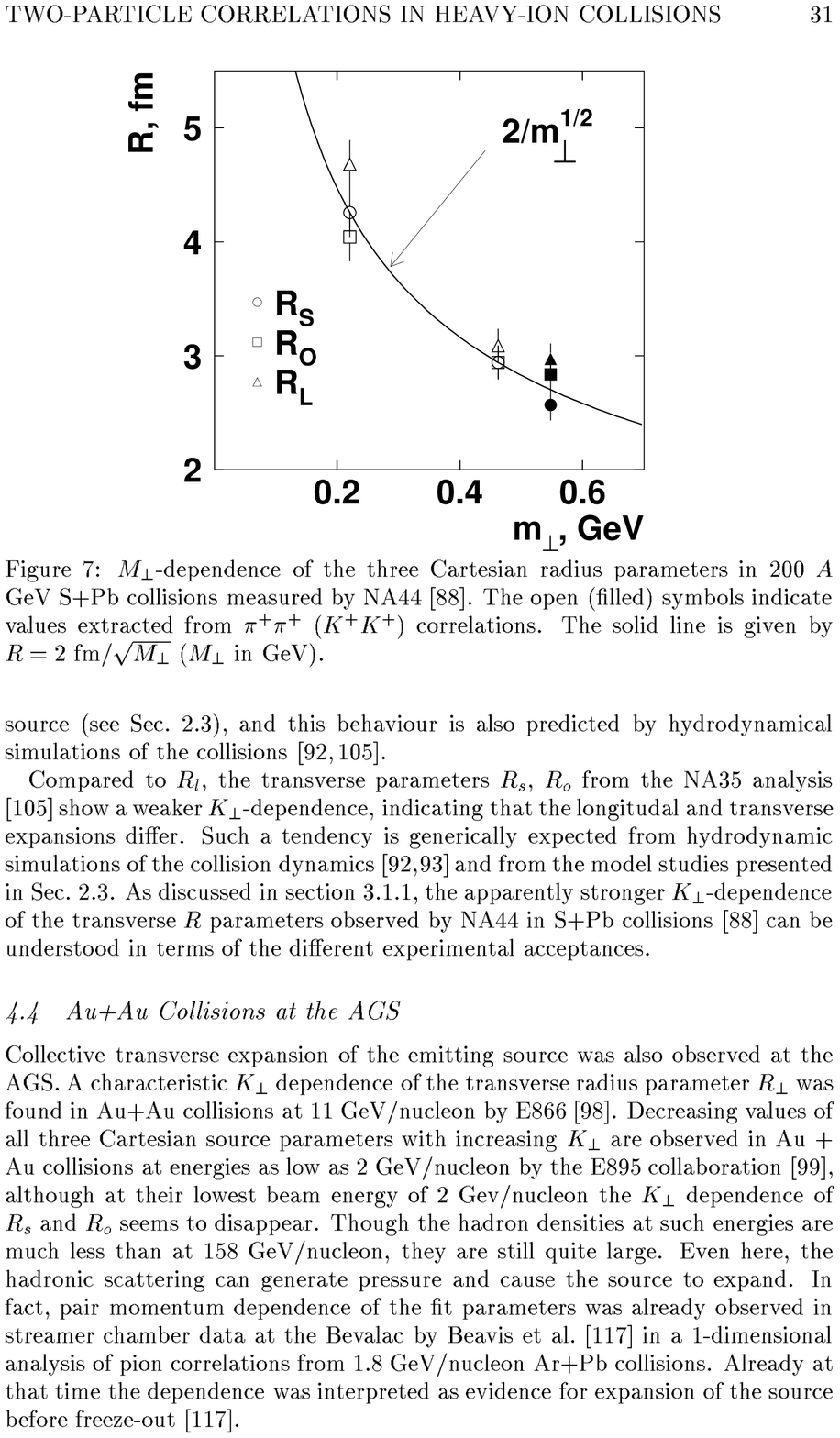,height=4.8cm,
bbllx=86pt,bblly=478pt,bburx=377pt,bbury=709pt,clip=}}
%bbllx=32pt,bblly=156pt,bburx=532pt,bbury=664pt,clip=}}
\caption{\small The $r$ as a function of $m_T$. 
Left:
DELPHI preliminary results from $e^+e^- \to Z^0 \to hadrons$. 
The line represents $r=0.354/\sqrt{m_T}$.
Right: NA44 results from
S+Pb collisions. The line represents $r=2/\sqrt{m_T}$. 
}
\label{fig_two}
\end{figure}

In recent years the correlations of pions emerging from
heavy ion collisions and from $e^+e^-$ annihilations have been analysed 
in terms of the 2-dimensional BEC analysis. 
This analysis is carried out in the Longitudinal Centre of Mass System,
the description of which can be found e.g. in Ref. \cite{review}.
In this system the direction of the event thrust is referred
to as the longitudinal direction and the two other directions 
are referred to as the outgoing 
and side axes. To these directions one associates three
four momentum differences $Q_z$, $Q_{out}$ and $Q_{side}$ which
in a 2-dimensional analysis are reduced to $Q_z$ and the transverse
momentum difference $Q_T=\sqrt{Q_{out}^2+Q_{side}^2}$.
In this method the emitter is not anymore restricted to a
spherical configuration but can also assume the shape of an
ellipsoid. Furthermore, one has an additional variable, 
the so called transverse mass $m_T$, defined as
$m_T=0.5\times(\sqrt{m^2+p_{T1}^2}+\sqrt{m^2+p_{T2}^2})$ 
where the $p_{Ti}$  
are the transverse momenta of the two hadrons.
From a fit to the data one is able to determine the longitudinal
$r_L$ and the transverse $r_T$ dimensions. 
The dependence of $r_L$ on $m_T$ are shown in Fig. \ref{fig_two}  
for $e^+e^-$ annihilations, as measured by DELPHI at 
the Z$^0$ mass, and
for S+Pb collisions at 200 GeV/A as measured by the NA44 collaboration. 
First to note is that $r_L(m_T)$ coming from $e^+e^-$
annihilations is very similar 
to $r(m)$ shown in Fig. \ref{fig_mass} which was obtained from the
1-dimensional BEC analyses of the Z$^0$ hadronic decay.
Moreover it has indeed been shown \cite{alex} that, as 
in the case of $r(m)$, also
$r_L(m_T)$ can be deduce from the Heisenberg uncertainty relations
to yield 
\begin{equation}
r_L(m_T)\approx c\sqrt{\Delta t ~\hbar}/\sqrt{m_T}
\end{equation}
which is the Eq. (\ref{eq_acl}) expression where $m$ and $r(m)$ are substituted
respectively by $m_T$ and $r_L(m_T)$. 
The dependence of $r_L$ on 1/$\sqrt{m_T}$ is repeated
also in the heavy ion collisions with the difference that the
proportionality factor is equal to 2.0 as compared to 0.354 found for 
$e^+e^-$ annihilations. To note is that the ratio
2.0/0.354 is not far from the ratio of the
average radius of the 
S and Pb nuclei to the radius of $\sim$0.85 fm obtained
from the 1-dimensional BEC of the hadronic Z$^0$ decays.
At this point the obvious question arises as to what extend one
is permitted
to extract from the 2-dimensional BEC analyses
information on the underlying dynamics of heavy ion collisions when
$r_L(m_T)$ behaves essentially the same as 
the in $e^+e^-$ annihilations
which are void of nuclear matter.
\section{Generalised Bose-Einstein Correlations (GBEC)}
In analogue to the 
generalised Pauli principle, which extends that principle
from two identical nucleons 
to the proton-neutron system,
one may consider an inclusions of the isospin invariance to extend
the BEC to a generalised BEC (GBEC).  
Such an extension          
has been proposed by several authors \cite{review} with the aim to
establish BEC relations between e.g. the $\pi^{\pm}\pi^{\pm}$
and $\pi^{\pm}\pi^0$ systems. One relation of this kind, which
was proposed by \cite{alex}, states that  
$$\sum_X\sqrt{ P[i_o \rightarrow (\pi^\pm\pi^\pm) X]} =
\sum_X\sqrt{ P[i_o \rightarrow (\pi^\pm\pi^0)_{even ~\ell} ~X]}$$
Here $P$ stands for probability, $i_0$ stands for an I=0 initial
state, like $\Upsilon\to b\bar b\to hadrons$,
and $X$ stands for the rest of the final state associated with the
specified two pion system. The GBEC validity at high energy
physics is still an open question due to the missing of an experimental
verification.
\section{Conclusions}
The $r(m)$ behaviour should be investigated also in other
than $e^+e^-$ particles reactions and hopefully the GBEC
applicability to hadron production will soon be tested experimentally.
In trying to gain an insight into the $r(m)$ behaviour 
and the 
energy density of the hadron emitter it may be instructive to turn
to other phenomena related to the Bose-Einstein
statistics.
One of them are the  Bose condensates where it was shown
\cite{alex} that at equal very low temperature 
their inter-atomic separations are proportional to 1/$\sqrt{m_{atom}}$
and not to their dimension. Moreover, in as much that the Heisenberg
relations are applicable to condensates, as 
most of their atoms are in the energy ground state,
the inter-atomic separation formula can be transformed 
to Eq. (\ref{eq_acl}) so that a comparison with  
the BEC dimension is not unreasonable. This comparison
coupled to the 
behaviour of $r(m)$ and the high emitter energy densities 
may point to the need to re-examine
the interpretation given to  
the BEC derived dimension $r$.

\end{document}